\documentstyle[aps,twocolumn]{revtex}
\begin{document}

{\bf Reply to Lambrecht, Jaekel, and Reynaud and to Garcia and Levanyuk:}

The two preceding Comments heavily criticize my recent suggestion that
the light emission in sonoluminescence might be explained in terms of
quantum vacuum radiation \cite{lett,pra}. I find that the criticism of
both of the Comments is unfounded. Lambrecht {\em et al} fallaciously
base general assertions on one particular model which is physically
ill-chosen, and Garcia and Levanyuk mistake two illustrational models
of Ref. \cite{pra} for realistic and make estimates for them under
inappropriate approximations. As neither of the two Comments is based
on an independent calculation but both rely entirely on formulae taken
from my papers \cite{lett,pra}, it appears to me that they have arisen
from an incomplete understanding of my theory. I will try and
elucidate a number of points.

Lambrecht {\em et al} estimate, in the short-wavelength limit, the
number of photons radiated by a spherical cavity whose time-dependent
radius $R(t)$ follows a Lorentzian model profile. I find no fault with
their back-of-the-envelope calculation, but I do not agree with their
rather general conclusion about fundamental limits of the amount of
radiation from a moving spherical cavity. The radiated energy and the
number of radiated photons are functionals of the radius $R(t)$ as a
function of time, and the properties of a functional can in general
not be derived by choosing just one particular kind of function as the
argument of the functional. In fact, all that Lambrecht {\em et al}
show is that a Lorentzian profile for $R(t)$ is not a physically
realistic choice, and this point has already been made in
Ref. \cite{pra} on p. 2780. Lambrecht {\em et al}'s conclusion that the
total number of radiated photons is fundamentally limited by the speed
of the bubble surface is proven erroneous by a simple counter-example
from Ref. \cite{pra}, pp. 2780--2781. The function $R(t)=R_{\rm min} +
\beta_{0}t\,\tanh(t/\gamma)$ leads to a radiated energy whose leading
term behaves like ${\cal W}\propto \beta_0^2/\gamma^3$. For this choice
of $R(t)$ the velocity $\dot{R}(t)=\beta(t)$ is bounded by
$1.2\beta_0$ but $\lim_{\gamma\rightarrow 0} {\cal W}$ diverges for
all non-zero $\beta_0$. Hence the radiated energy is not bounded from
above, QED. Ditto for the number of radiated photons. As discussed in
Refs. \cite{lett,pra} (cf. e.g. the paragraph below eq. (10) of
\cite{lett}) the physical reason for this is that the amount of quantum
vacuum radiation is to leading order governed not by the velocity
$\beta(t)$ nor by the acceleration $\dot{\beta}(t)$ but by the fourth
time-derivative $\beta^{(4)}(t)$ of the velocity. Lambrecht {\em et
al}'s problems arise because Lorentzians form a one-parameter family
of functions, and hence the velocity $\beta$ and its fourth derivative
$\beta^{(4)}$ are necessarily governed by the same one parameter.

Turning to the Comment by Garcia and Levanyuk, I should like to point
out that although the above argument illuminates some principal
properties of quantum vacuum radiation it can hardly be used for
reliable estimates of the intensity. Apart from the fact that, as
illustrated by the above discussion, the result of any such estimate
does indeed depend very strongly on the particular choice of a model
function for $R(t)$, both Lambrecht {\em et al} and Garcia and
Levanyuk seem to have overlooked the fact that all the equations for
the photon number and the radiated energy they quote contain
substantial approximations, most notably a short-wavelength expansion
(cf. fn. [18] of Ref. \cite{lett} and the extensive discussion on
pp. 2779--2782 of Ref. \cite{pra}), which is to say that these
expressions are valid only if the wavelength of the emitted light is
very much shorter than the size of the cavity during emission --- a
condition that is not satisfied for sonoluminescent bubbles. When the
wavelength and the cavity radius are of the same order of magnitude,
resonance effects occur, as is well-known from scattering theory
(cf. e.g. \cite{hulst}). For a dielectric sphere in an optically
thinner medium such resonances lie at real arguments $kR$ of the
spherical Bessel functions and lead to the widely known
whispering-gallery modes. In the converse case of a spherical cavity
in an optically denser medium these resonances occur for complex
arguments of the Bessel functions and hence are more difficult to keep
track of. As an additional complication the case of quantum vacuum
radiation from a sphere brings about products of four spherical Bessel
functions (cf. eq. (4.3) of \cite{pra}) and not just two as in the
standard Mie theory of light scattered from spheres, so that no known
analytical techniques can be resorted to. That is why the spectrum of
quantum vacuum radiation was calculated numerically in
Ref. \cite{pra}. For comparison of the numerical results with those
obtained analytically in the short-wavelength limit, a Lorentzian
model function for $R(t)$ was introduced. The reason for choosing a
Lorentzian was that it is governed by just one parameter and its
Fourier transform is a pure exponential which forestalls the need for
any asymptotic approximations in the analytical calculations. The
comparison showed that resonance effects lead to a substantial
enhancement over the estimates made in the short-wavelengths
approximation.

The purpose of the second model function introduced in Ref. \cite{pra}
and picked up by Garcia and Levanyuk in their eq. (4) was then just to
show explicitly, in the short-wavelength limit, that the amount of
quantum vacuum radiation is not principally limited by any limit to
the velocity, as discussed above, and that the superluminal velocities
arising for Lorentzian model functions are an artefact of a particular
choice of model function but not of any physical meaning. The
attempts by Garcia and Levanyuk of fitting this model function to
experimental data may be correct but bear no significance to any
general estimate of the amount of quantum vacuum radiation from a
collapsing gas bubble in a fluid.

The difficulty of finding a realistic model for the dynamics of the
cavity radius $R(t)$ is twofold. First, there is no good way of
estimating the magnitude of the fourth time-derivative of the velocity
from general physical arguments; and second, the system of a moving
spherical cavity is subject to a back-reaction from the emitted light,
i.e. one is dealing with a coupled and highly nonlinear system, as
has been emphasized before (cf. Sec. V.C of Ref. \cite{pra}).

Neither Lambrecht {\em et al} nor Garcia and Levanyuk seem to be
sufficiently familiar with the phenomenon of sonoluminescence 
to have realized that the crucial part of the dynamics of a collapsing
sonoluminescent bubble, namely the dynamics in the vicinity of the
collapse, is not described by the Rayleigh-Plesset equation. Indeed
much current research effort about sonoluminescence is directed
towards the accurate description of the bubble dynamics. The works
cited by Lambrecht {\em et al} and by Garcia and Levanyuk emphasize
that the Rayleigh-Plesset equation is valid only for low Mach numbers
and in the absence of any shock waves and that it therefore can serve
at best as a crude approximation to the rather complicated problem of
a collapsing sonoluminescent bubble. Extensive numerical simulations
of the dynamics inclusive of shock waves have been performed by several
authors \cite{wu,moss,kondic,szeri}. In particular, Vuong's and
Szeri's recent calculations have shown very clearly that the dynamics
of the bubble radius $R(t)$ close to its minimum must be expected to
involve at least picosecond timescales \cite{szeri}. In view of the
discussion on pp. 2781 and 2782 of Ref. \cite{pra} this raises hopes
on quantum vacuum radiation as a possible candidate for the
explanation of the light emitting mechanism.

As regards the experimental technique of measuring the time-dependence
of the bubble radius by Mie scattering from the bubble, the authors of
the preceding Comments are apparently unaware that these experiments
are fundamentally limited to at least several nanoseconds in their
time resolution, although this is pointed out in the papers cited by
Lambrecht {\em et al} and by Garcia and Levanyuk. One reason for this
limited resolution is that the photomultiplier tubes have a finite
rise time; the other is that the data collection takes place over at
least several thousands of acoustic cycles so that the jitter in the
bubble dynamics washes out any fast timescales in the recorded data.
Hence one cannot expect to see picosecond or even femtosecond
timescales in these experiments which therefore cannot be cited as
proof of the absence of such timescales.

Finally, when it comes to estimating the fastest time\-scale in the
system, I do not think I agree with Garcia and Levanyuk. At least, I
do not know what sound should be at interatomic distances. If $10^6$
photons are radiated by a bubble of $0.5\mu$m radius then this means
that only 1\% of the water molecules in the top layer of the boundary
send off a photon each on average. This being the case, the shortest
timescale in the system with regard to quantum vacuum radiation is
that of the dynamics of this innermost layer of water molecules and
this is given by the interparticle collision time of the water
molecules with the gas molecules inside the bubble. The interparticle
collision time depends very strongly on density and local pressure and
under the given circumstances it can certainly lie in the
subpicosecond range. Refs. \cite{lett,pra} have argued that quantum
vacuum radiation could presumably still be a viable explanation for
the light emission in sonoluminescence if the shortest timescale in
the dynamics of the bubble surface is as long as 100fs, because the
crude estimate in eq. (9) of \cite{lett} does not account for the
resonant enhancement away from the short-wavelength limit.

In summary, I find that although both of the preceding Comments are
expressed with much emphasis they are based on inconclusive
arguments. On the basis of the present experimental evidence there is
no reason to eliminate quantum vacuum radiation as a possible
candidate for the explanation of the light observed in
sonoluminescence.

I would like to thank Andrew J. Szeri for sending me an advance copy
of his paper \cite{szeri}. Financial support through the Ruth Holt
Research Fellowship at Newnham College Cambridge is gratefully
acknowledged.

\vskip 0.15 in
\noindent
Claudia Eberlein\\
TCM, Cavendish Laboratory, University of Cambridge\\
Cambridge CB3 0HE, England.

\end{document}